\documentclass[]{elsarticle}

\usepackage{lineno,hyperref}
\usepackage{amsmath}
\usepackage[justification=centering]{caption}
\usepackage[autostyle=once]{csquotes}

\usepackage[bottom]{footmisc}
\usepackage{graphicx}
\usepackage{tabularx}
\usepackage{xcolor}
\usepackage{xspace}
\usepackage{verbatim}

\newcommand{\concept}[1]{\texttt{#1}}
\newcommand{\takeaway}[2]{\noindent\fbox{\parbox{\textwidth}{\textbf{#1}{#2}}}}

\modulolinenumbers[5]

\journal{arXiv}

\begin{document}

\begin{frontmatter}

\title{Google Summer of Code: Student Motivations and Contributions}

\author{Jefferson O. Silva, Igor Wiese, Daniel M. German, Christoph Treude, Marco A. Gerosa, Igor Steinmacher}

\begin{abstract}

Several open source software (OSS) projects expect to foster newcomers' onboarding and to receive contributions by participating in engagement programs, like Summers of Code. However, there is little empirical evidence showing why students join such programs. In this paper, we study the well-established Google Summer of Code (GSoC), which is a 3-month OSS engagement program that offers stipends and mentors to students willing to contribute to OSS projects. We combined a survey (students and mentors) and interviews (students) to understand what motivates students to enter GSoC. Our results show that students enter GSoC for an enriching experience, not necessarily to become frequent contributors. Our data suggest that, while the stipends are an important motivator, the students participate for work experience and the ability to attach the name of the supporting organization to their resum\'{e}s. We also discuss practical implications for students, mentors, OSS projects, and Summer of Code programs.

\end{abstract}

\begin{keyword}
Google Summer of Code \sep Motivation \sep Newcomers \sep Open Source Software
\MSC[2018] 00-01\sep  99-00
\end{keyword}

\end{frontmatter}



\section{Introduction}
\label{intro}

Summer of Code programs aim at promoting software development by students for a few months ~\cite{Silva2017c, Trainer2014b}. By participating in these programs, Open Source Software (OSS) projects expect to increase newcomers' retention and code contribution~\cite{Trainer2014b}. Examples of such programs include Google Summer of Code,\footnote{\href{https://developers.google.com/open-source/gsoc/}{ http://developers.google.com/open-source/gsoc/}} Rails Girls Summer of Code,\footnote{\href{http://railsgirlssummerofcode.org/}{ http://railsgirlssummerofcode.org/}} Julia Summer of Code,\footnote{\href{https://julialang.org/soc/archive.html}{ https://julialang.org/soc/archive.html}} and Outreachy.\footnote{\href{https://www.outreachy.org/}{ http://www.outreachy.org/}} Some Summer of Code programs are sponsored by well-known organizations, such as Facebook, Yahoo!, and Google ~\cite{Trainer2014b,Trainer2014c}.
Nevertheless, students that participate in Summer of Code programs are likely to have personal goals beyond becoming active OSS project contributors, such as building their CV or receiving stipends ~\cite{Tirole2002, Lakhani2005}.

Previous research has mostly focused on new ways to attract developers into OSS (e.g., \cite{Meirelles2010,Santos2013}), to retain them as long-term contributors (e.g., \cite{VonKrogh2003,Fang2009b,Ducheneaut2005}), and to mitigate onboarding barriers (e.g., \cite{Steinmacher2015}). Regarding Summer of Code programs, the literature has focused on quantitative evaluations of the contributions made by the students during and after the programs~\cite{Schilling2011g} (for a few projects of the KDE community); and on the outcomes for the students that participated in these programs ~\cite{Trainer2014b, Trainer2014c, Trainer2016a}.
No research has focused on the motivations that these students had to join an OSS project and the influence that being part of the program (such as the gain in reputation and the pecuniary benefits of joining the program) has on their motivations; neither has research explored the perspective that mentors (members of the OSS projects) have about the students' motivation.

Thus, the purpose of this study is to identify and understand what motivates students to participate in Google Summer of Code (GSoC) programs and to continue participating in the projects after the program end. We chose to focus our study on GSoC because it is the oldest, largest, and best-known Summer of Code program. We collected data by means of surveys and interviews with students and mentors in order to promote triangulation of data sources.

We designed the following research questions (RQ) to guide our research:

\paragraph{\textbf{RQ1}} According to students, what motivates them to participate in Summer of Code programs?

\paragraph{\textbf{RQ2}} According to mentors, what motivates students to participate in Summer of Code programs?

Our findings suggest that most students participate in Summer of Code programs to acquire experiences and technical skills that can be used later for career building. Nevertheless, for a small number of students, their desire to contribute to an OSS project---even after the programs---is more than a participation bonus, but an experience they do not want to forgo. We conjecture that OSS projects could increase the odds of achieving students' retention by providing the students with participation rewards (e.g., certificates) aligned with the students' interests (e.g., career building).


\section{Background and Related Work}
\label{sec:related-work-background}

In this section, we summarize studies that tackled not only the newcomers' self-guided involvement in OSS projects but also their involvement through Summers of Code. We start by explaining what Google Summer of Code is, how it works, and why we chose to study it.

\subsection{Google Summer of Code}

Google Summer of Code (GSoC) is a worldwide annual program sponsored by Google that offers students a stipend to write code for OSS for three months. We chose to study GSoC because it is best-known compared to other programs; has been in operation since 2005; every year a large number of students from all over the world participate in it, and it provides students with a comprehensive set of rewards, including participating in a well-known large company's program, community bonding, skill development, personal enjoyment, career advancement, peer recognition, status, and a stipend~\cite{Trainer2014b}.

Among its goals, GSoC aims to "Inspire young developers to begin participating in OSS development," and "Help OSS projects identify and bring in new developers and committers."
\footnote{\href{https://google.github.io/gsocguides/student/}{ https://google.github.io/gsocguides/student/}}
At the time of this writing, Google paid 3,000 to 6,600 USD (depending on the country) for students who successfully complete all phases of the program.

Applicants must write and submit project proposals to the OSS projects (previously approved by Google) they wish to work for.
Accepted students spend a month learning about the organization's community and, then, three months implementing their contribution, which is evaluated by the mentors before they receive the final payment.

\subsection{Summer of Code Programs}

Summer of Code programs are becoming a common initiative to bring more contributors to OSS (e.g.,Google Summer of Code, Julia Summer of Code), and to increase diversity (e.g., Outreachy, Rails Girls Summer of Code). 
Given Summer of Code aparent success, some researchers have targeted  these programs to understand students' retention. For example, Schilling et al.~\cite{Schilling2011g, ScLW11} used the concepts of Person-Job (the congruence between an applicant's desire and job supplies) and Person-Team (the applicant's level of interpersonal compatibility with the existing team) from the recruitment literature. They found that intermediate (4-94 commits) and high (\textgreater94 commits) levels of previous development were strongly associated with retention. Trainer et al.~\cite{Trainer2014c} interviewed 15 students and identified the students gained new software engineering skills, and the students used their participation for career advancement. The authors also found that mentors faced several challenges. In another study, Trainer et al.~\cite{Trainer2014b} analyzed 22 GSoC projects in the scientific software domain to understand GSoC outcomes. They found that GSoC facilitated the creation of strong ties between mentors and students, reporting that 18\% of the students (n=22) became mentors in subsequent editions.


\subsection{Motivation}

A conventional understanding among researchers seems to be that motivation refers to the psychological needs that require satisfaction~\cite{Deci1999a}. These needs can be acquired through the influence of the environment or they can be innate~\cite{Mason2012}. As with other practitioners, software engineers are influenced by their motivational state, which is determinant to the success or failure of software projects~\cite{Beecham2008b}.

We focus on the OSS context, and it is out of the scope of this study to provide an exhaustive systematic review of motivational theories. Instead, we chose to study students' motivation using the constructs of intrinsic and extrinsic motivation and the self-determination theory, which have been frequently used to analyze OSS project developers (see \cite{BeBe10} and \cite{VonKrogh2012} for a review).


Intrinsically motivated behaviors do not require any 'rewards' other than those obtained from the satisfaction of performing them~\cite{Deci1999a}. In contrast, extrinsically motivated behaviors are the pursuit of external rewards or the consequences derived from their performance~\cite{SDPR92}. Individuals can undergo a motivation internalization process that moves 'pure' extrinsic motivations closer to 'pure' intrinsic motivations, considering that motivation is a continuum, which is referred to as internalization of extrinsic motivations~\cite{Roberts2006a}.


The Self-Determination Theory (SDT) is a general motivational theory, which is concerned with motivation behind individual choices~\cite{Deci1999a}. Several researchers built upon SDT to explain the heterogeneous nature of individual's motivation in a broad range of domains~\cite{BeBe10,Deci1999a}, including OSS developers' motivation to contribute voluntarily to OSS projects. For example, several empirical studies found intrinsic motivation factors that played a significant role in motivating OSS developers, such as:
\textit{ideology}~\cite{Lakhani2005,Ghosh2005a}
\textit{altruism}~\cite{Ghosh2005a,BiSS07,HaPS03};
\textit{kinship amidity}~\cite{Lakhani2005,David2008};
and \textit{enjoyment and fun}~\cite{Shah2006,Lakhani2005}
Several internalized extrinsic motivation factors were found to be important, such as
\textit{reputation}~\cite{Ghosh2005a, Spaeth2008,Lakhani2003};
\textit{reciprocity}~\cite{Lakhani2005,Lakhani2003};
\textit{learning}~\cite{Ghosh2005a,Spaeth2008,Hippel2003}; and
\textit{own use value}~\cite{Lakhani2005,Ghosh2005a,HaOu02}. We highlight that the most commonly cited extrinsic motivation factors are \textit{career building}~\cite{Tirole2002, HaOu02} and \textit{stipends}~\cite{Lakhani2005,HaOu02,Litheger2007}.

\subsection{Newcomers' Onboarding}

Typically, studies on retention take the perspective of the individual developer. Thereby, intrinsic motivation (e.g.,~\cite{Lakhani2005,HaOu02}), \textit{social ties} with team members (e.g.,~\cite{Fagerholm2014c,Steinmacher2015c,Steinmacher2014}), \textit{mentoring} (e.g., \cite{ScLW12}), \textit{project characteristics} (e.g.,~\cite{Santos2013,Colazo2009,Meirelles2010}), \textit{ideology} (e.g.,~\cite{Stewart2006}), and \textit{incentives} and \textit{rewards} (e.g.,~\cite{Hann2002b,Krishnamurthy2014a}) have been found most relevant for OSS developers to remain contributing.

Zhou and Mockus~\cite{ZhMo12} worked on identifying newcomers who are more likely to remain contributing. They found that the \textit{individual's willingness} and the \textit{project's climate} were associated with the odds that an individual would become a long-term contributor. Similarly, Wang and colleagues \cite{WaWa18} proposed a prediction model to measure the chance for an OSS software developer become a long-term contributor. The authors found that \textit{willingness} and the \textit{environment} were associate with newcomers becoming long-term contributors.

Fang and Neufeld~\cite{Fang2009b} built upon the Legitimate Peripheral Participation (LPP) theory ~\cite{Lave1991a} to understand developers' motivation. Results from qualitative analyses revealed that \textit{initial conditions to participate} did not adequately predict long-term participation, but that \textit{situated learning} and \textit{identity construction} behaviors were positively linked to sustained participation. From another perspective (including LPP lens), Sholler et al.~\cite{Sholler.ea:2019} built upon existing literature to provide rules for helping newcomers become contributors to OSS projects.


\section{Research Method}
\label{sec:research-method}

\begin{figure}[hbtp]
  \centering
  \includegraphics[width=\textwidth]{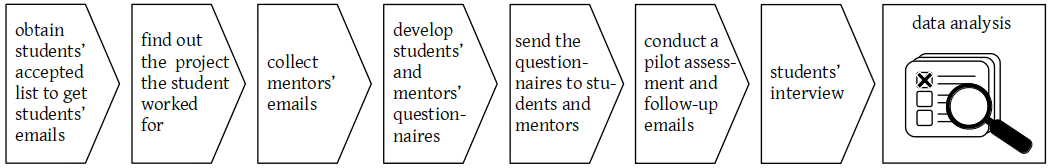}
  \caption{Research Method}
  \label{fig:research_method}
\end{figure}

To answer our RQs, we conducted surveys with students and mentors and follow-up interviews with students. Figure~\ref{fig:research_method} outlines the research method we followed in this study. We conducted surveys not only to assess the motivational factors we found in the current literature but also to uncover potential new ones.

\subsection{Contact information collection}

The first step of our study was to search for information (e.g., email addresses) that would enable us to contact the students. We used the accepted students' list, published by Google, which contains the students' and the OSS organizations' names. Based on this information, we investigated which specific project a student worked for, considering all the OSS projects under each organization. For example, although Google informs that the Apache Software Foundation (organization) accepted participant John Doe, we still do not know for which Apache project John worked. We considered that we found their emails when we had clear evidence linking the student with their corresponding project name. For instance, when we found students' web blog or their professional resum\'{e}s describing their experience in the program, or when we found their messages about the program in projects' discussion lists.

As the collection and verification of each student project is laborious and time-consuming, we limited our analysis to the GSoC 2010-2015 editions, in which approximately 7,000 students participated.\footnote{\href{https://developers.google.com/open-source/gsoc/resources/stats}{http://developers.google.com/open-source/gsoc/resources/stats}} By the end of this step, we could gather 1,000 students' and 730 mentors' emails.

\subsubsection{Questionnaire design and administration}
We used questionnaires as a data collection method, following Fink's advice on how to design surveys~\cite{Fink1995}. We asked students\footnote{The students' questionnaire is available at \href{https://docs.google.com/forms/d/e/1FAIpQLSfqLGFz3KdTiiD43s4tnxOaOy0vOPjd2vYuoq38uycYosLcRQ/viewform}{http://docs.google.com/forms/students}} about their contributions to OSS before and after GSoC (questions 1-5) and general questions about their participation in GSoC (questions 6-13). We also asked them questions that further explored the relationship between stipends and participation in GSoC (questions 14-15) and whether they would enter a hypothetical-GSoC that offered all motivational factors but one (question 16), which allowed us to rank and examine how essential these factors were. We concluded by asking them about demographic information at the time of their first participation (questions 17-22).

We designed the mentors' questionnaire\footnote{The mentors' questionnaire can be accessed at \href{https://docs.google.com/forms/d/e/1FAIpQLSd-J9EJKnmZSJ5CsjMuvSDeGgzbC7wb7-WctLnKZx7aGcAq4A/viewform}{http://docs.google.com/forms/mentors}} using the same structure as the students', with the difference that mentors had to answer about their students in general. It is worth emphasizing that we are aware that the mentors' answers may not refer to the students in our sample but they can provide a more complementary point of view.

We conducted a pilot assessment of the questionnaire with 2 GSoC 2015 students. After minor adjustments, we sent out emails inviting students to participate in this research. We employed principles for increasing survey participation~\cite{SmLM13}, such as sending personalized invitations, allowing participants to remain anonymous and sending follow up emails.

We sent out 1,000 survey invitations ($\approx$14\% of the total GSoC students for the investigated period) to students and received answers from 141 students (14.1\% response rate). We also sent out 730 survey invitations to mentors, and we received 53 responses (7.3\% response rate). The number of survey invitations sent out to mentors is smaller than that of the students because a considerable number of mentors participate in more than one edition.

\subsection{Analysis of survey responses}

We employed descriptive statistics for analyzing the answers to the closed-ended questions and open coding and axial coding~\cite{Strauss1998} for the open-ended ones. Open coding involves identifying codes and their properties in the data. Axial coding involves relating data together in order to reveal concepts and categories via a combination of inductive and deductive thinking \cite{Cres12}.

The first author performed the open coding in the first stage, which resulted in 481 different codes. Two other authors collaborated to derive the 17 concepts from these codes. In the second stage, a third author reviewed the concepts and collaborated in the generation of the 7 categories, as presented in Table~\ref{tab:motivation_concepts}.

In the findings section, we provide a selection of representative quotes from students and mentors, denoted respectively by S\textsubscript{\#}, and M\textsubscript{\#}, with their IDs in subscript. We also present in parentheses how many participants mentioned a category or concept. The counts represent how much evidence the data analysis yielded for each theme; they do not necessarily mean the importance of a theme.

\subsection{Semi-Structured Interviews}

We interviewed the surveyed students who volunteered for follow-up online interviews to enlighten some motivation factors that were still unclear. Besides, we wanted to get their perception of the coding scheme we derived during the survey analysis. We crafted the interview questions following Merriam's~\cite{Merriam2009} advice, to stimulate responses from the interviewees.

We sent out 43 invitation emails and received 10 positive responses. The interviews lasted, on average, 23 minutes. At the end of the interviews, we presented and explained our coding scheme derived from the survey analysis, and asked for changes or insights that the students might have. Two interviewees suggested minor changes, such as including buying hardware equipment for participation as one of the roles of the stipends.

\subsection{Sample Characterization}

Our sample comprises 112 male students, two females, and two self-identified as other. The predominant age for the first participation in GSoC was between 21-25 years old (63), followed by 18-20 years old (45). A minority of students were between 26-30 years old (26) and 31-40 years old (7). Regarding education, the respondents were mostly undergraduate students (58) or held a bachelor degree (41) students.

A smaller number of students were graduate students (7) or held a graduate degree (6).
Most participants had previous development experience ranging from 2-4 years (62), and 5-9 years (41).

In comparison, GSoC published statistics on students' demographics for GSoC 2014\footnote{\href{https://opensource.googleblog.com/2014/06/google-summer-of-code-2014-by-numbers.html}{ https://opensource.googleblog.com/2014/06/gsoc-2014-by-numbers.html}} (we could not find other years' detailed statistics). For that year, 10\% of the students were females, $\approx$68\% of them were undergraduates, and they were typically between 18-25 years old. Our sample resembles these characteristics.

We also analyzed the students' distribution per country, shown in Table~\ref{tab:sts_by_countries}. We received answers from participants from 34 countries. Approximately 23\% of the students resided in India and $\approx$15\% of them in the USA. In comparison with GSoC published statistics from 2013,\footnote{\href{https://opensource.googleblog.com/2013/06/google-summer-of-code-2013-full-of.html}{ https://opensource.googleblog.com/2013/06/gsoc-2013-full-of.html}} 2014,\footnote{\href{https://opensource.googleblog.com/2014/05/google-summer-of-code-2014-by-numbers.html}{ https://opensource.googleblog.com/2014/05/gsoc-2014-by-numbers.html}} and 2015,\footnote{\href{https://opensource.googleblog.com/2015/05/gsoc-2015-stats-part-1-all-about.html}{ https://opensource.googleblog.com/2015/05/gsoc-2015-stats-about.html}} the sample is also representative regarding country.

\begin{table}
  \caption{Students' count per country of residence at the time of first participation}
  \label{tab:sts_by_countries}
  \begin{tabularx}{\textwidth}{X}
    \includegraphics[trim=2cm 15.5cm 2cm 2.5cm, clip, width=\textwidth]{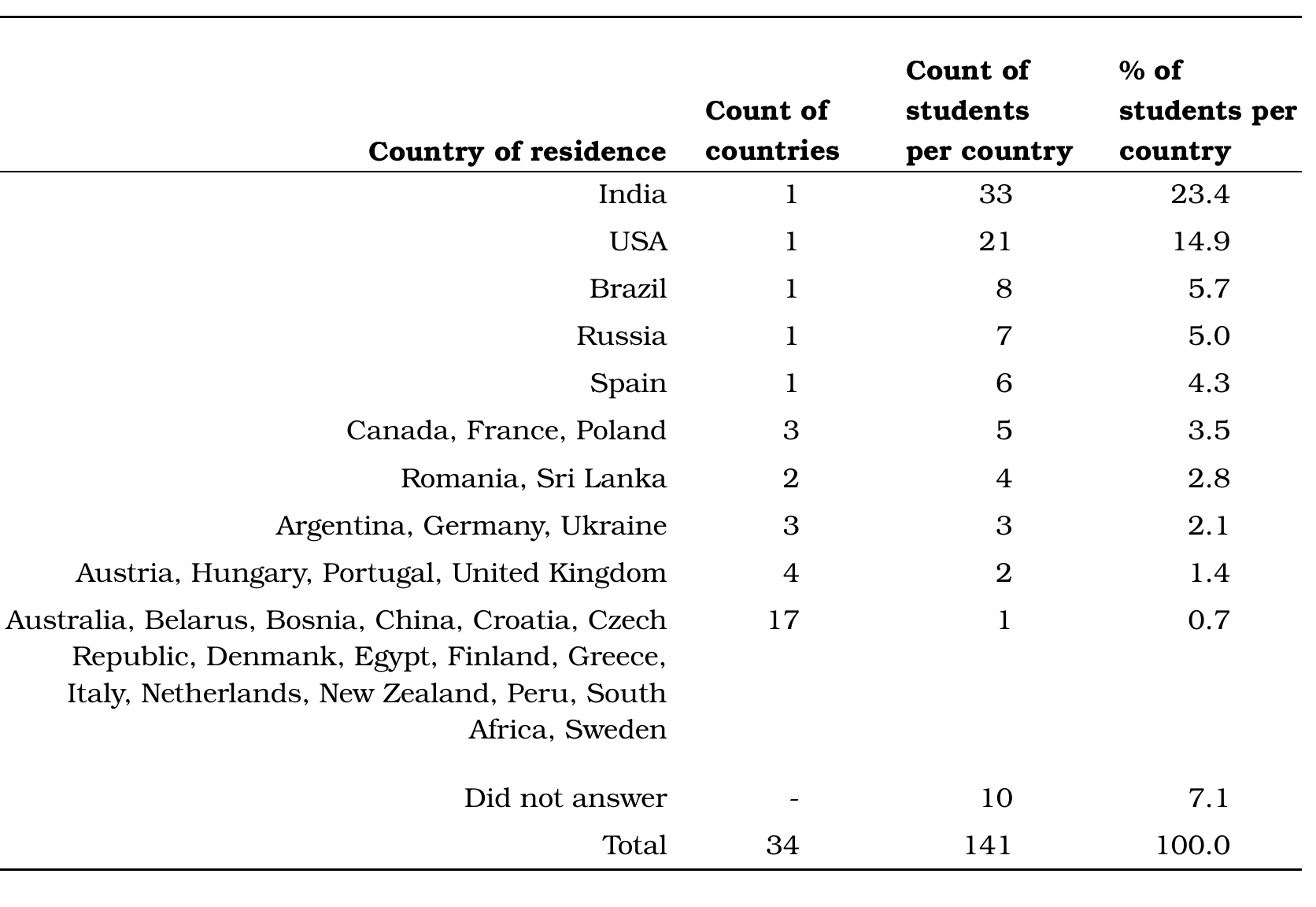}
  \end{tabularx}
\end{table}

\subsubsection{Demographic information about mentors}

All respondent mentors identified as males (53). Half of them were between 31-40  years old (27), 15 were more than 40, 10 were between 26-30, and only one was between 21-25. The respondents participated (as mentors) in: 1 edition (10); 2 editions (15); 3 editions (13); 5 editions (11); 6 editions (2); 7 editions (1); and 11 editions (1). Most mentors had more than ten years (44) of development experience, with a few that had seven years (5), six years (2), five years (1), and eight years (1).


\section{Findings}
\label{sec:findings}

\subsection{Students' Motivations to Join GSoC (RQ1)}

\begin{figure}[htb]
  \centering
  \includegraphics[scale=1.1]{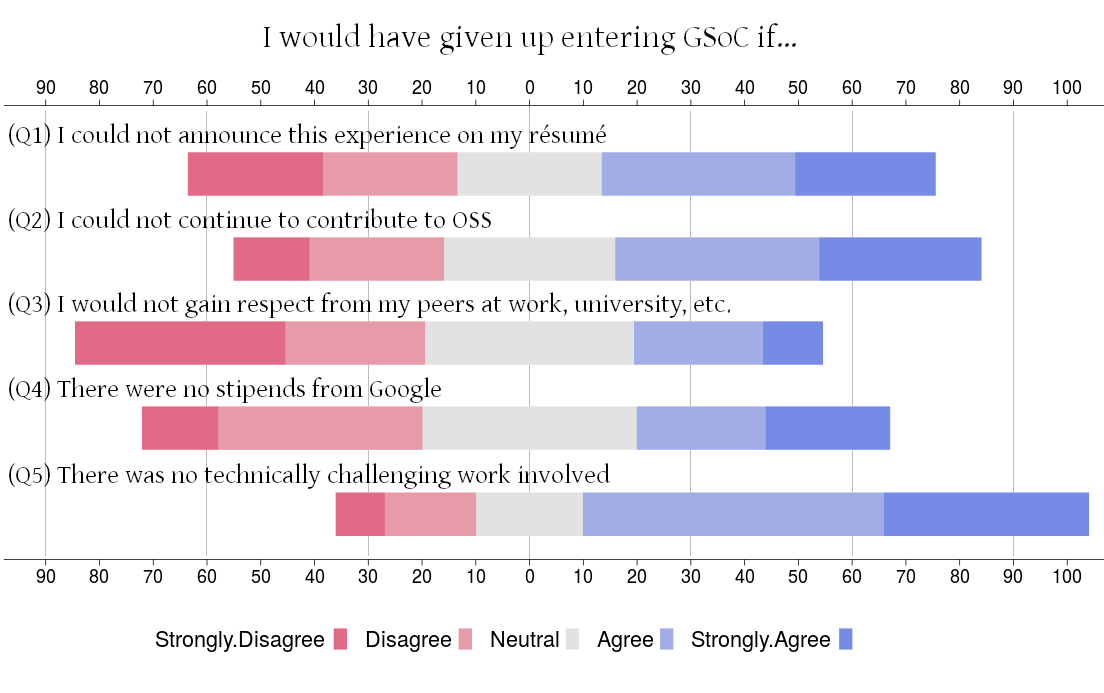}
  \caption{Students' assessment of motivation factors for participating in GSoC}
  \label{fig:sts:motivators_assessment}
\end{figure}

Based on the literature (e.g.,~\cite{Beecham2008b}), we asked how essential the following motivation factors were for the students to participate in a hypothetical-GSoC that offered all factors but one: career building (Q1); an entry gateway to OSS projects (Q2); peer recognition (Q3); stipends (Q4); and intellectual stimulation, such as a technical challenge (Q5). Figure~\ref{fig:sts:motivators_assessment} depicts in stacked bars the agreement level (5-level Likert items). We considered a motivation factor essential when the students reported they would give up entering the hypothetical-GSoC without that factor.

\begin{figure}[htb]
  \includegraphics[trim=2cm 18.5cm 2.9cm 2.4cm, clip, width=\textwidth]{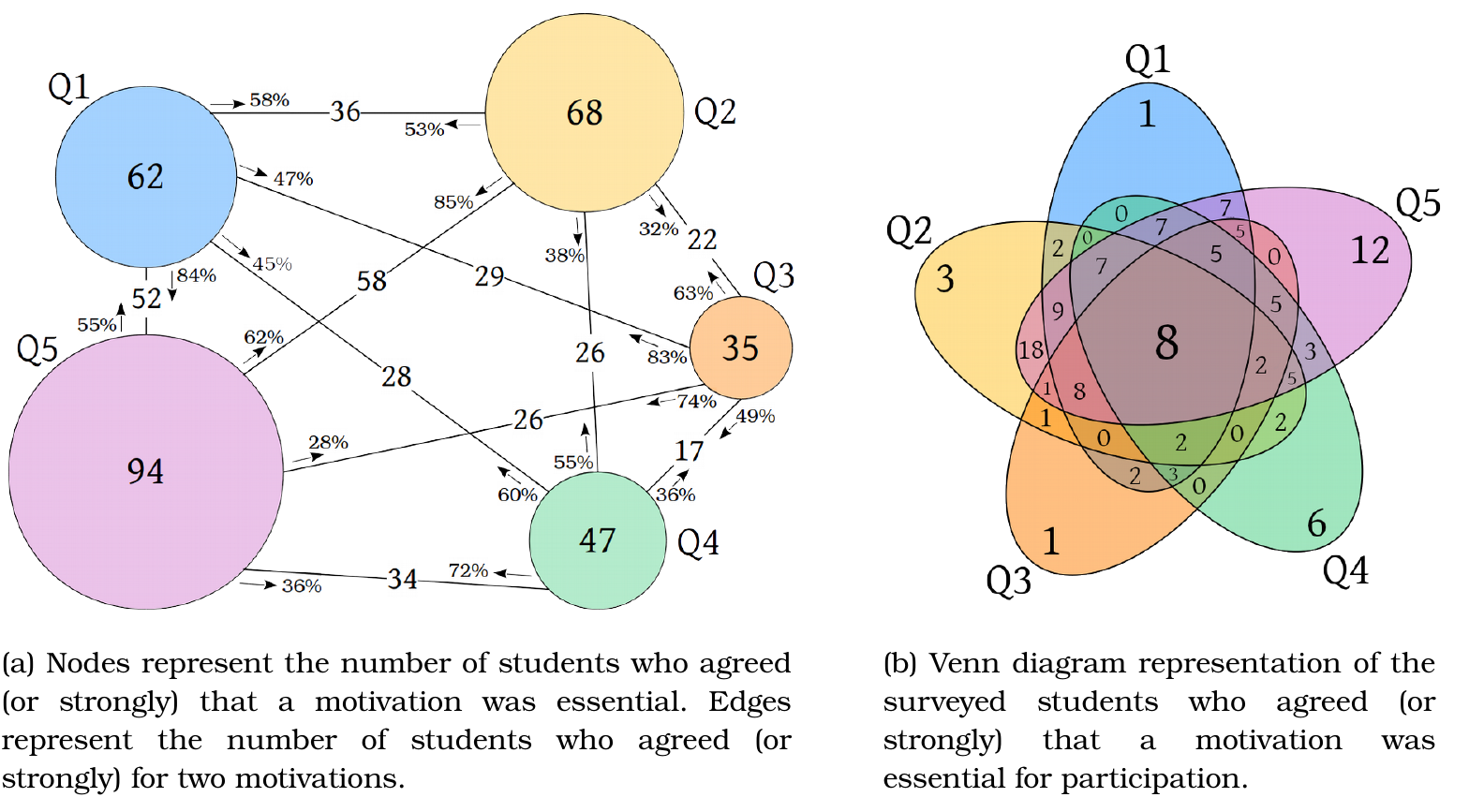}
  \caption{Surveyed students' motivation count in a graph (a) and in a Venn diagram (b). Career building (Q1); contribute to OSS (Q2); peer recognition (Q3); stipends (Q4); technical challenge (Q5) }
  \label{fig:sts:graph-venn-141-agreers}
\end{figure}

In Figure~\ref{fig:sts:graph-venn-141-agreers} (a), we offer an alternative perspective, with the students' responses presented in a graph, highlighting counts, proportions, and how the motivations factors relate to each other in pairs. Each node in this figure indicates the number of students who considered that factor essential. Node sizes are proportional to the counts. The edges depict the counts in the intersection of two motivation factors. Percentages show the proportion of the intersection in relation to a node (i.e., motivation factor). In Figure~\ref{fig:sts:graph-venn-141-agreers} (b), we decompose the students' response counts into sets and subsets, with the results shown in a Venn diagram.

The analysis of students' textual answers yielded motivation factors other than the ones that triggered our investigation, such as \textit{learning} and \textit{academic} concerns. Table~\ref{tab:motivation_concepts} presents all the concepts and categories derived from the students' answers.

\begin{table}
  \centering
  \caption{What motivates students to participate in Google Summer of Code?}
  \label{tab:motivation_concepts}
  \begin{tabular}{c}
    \includegraphics[trim=2.5cm 5.3cm 2.2cm 2.7cm, clip, scale=0.6]{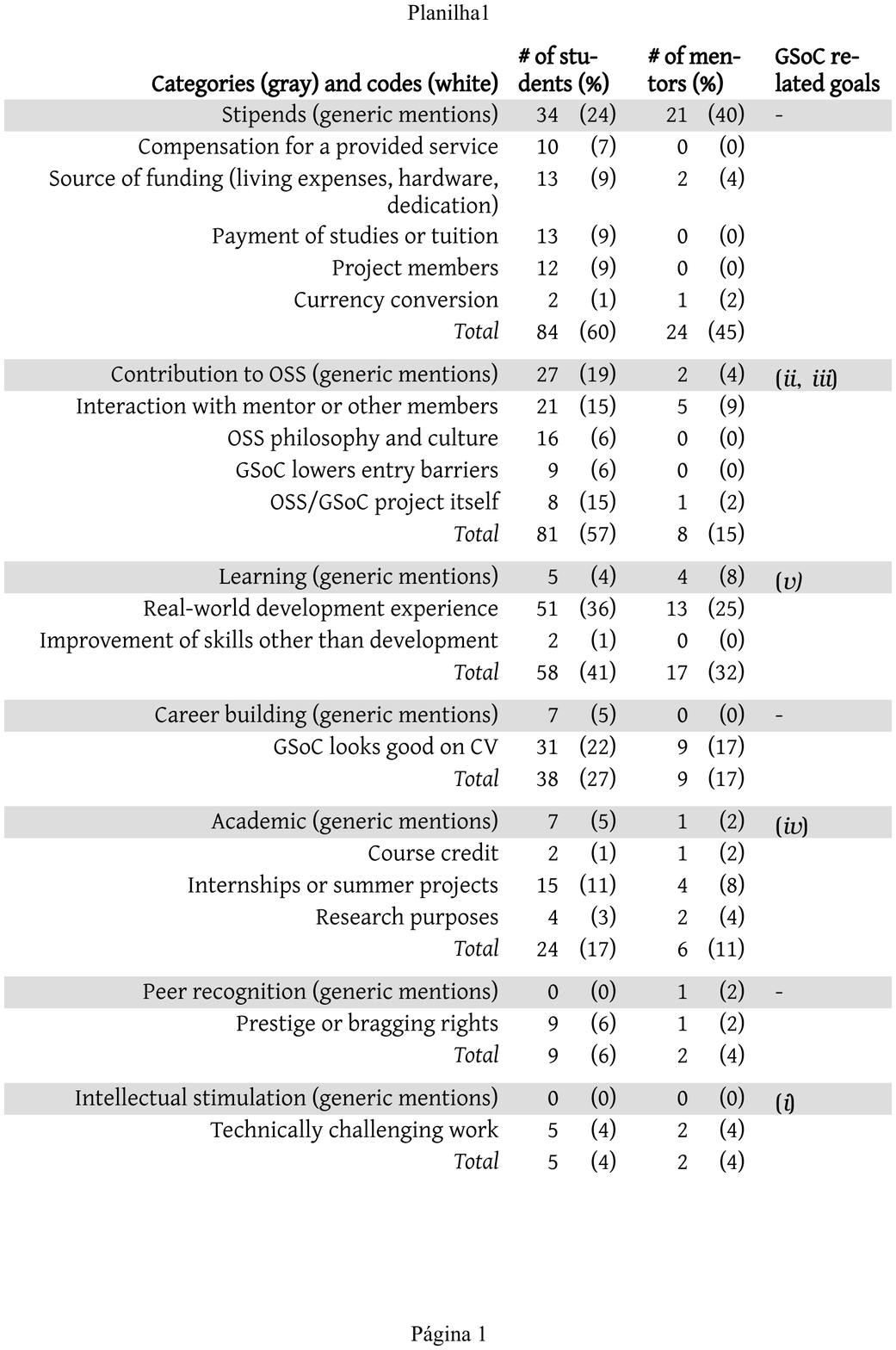}
  \end{tabular}
\end{table}

For readability concerns, we adopt the following convention to present the results in Table~\ref{tab:motivation_concepts}. Concepts are presented in True Type font (\concept{concept}) (1). Categories are presented in italics (\textit{category}) (1). Totals are presented in boldface (\textbf{total}) (1). In all cases, the numbers in parentheses depict the counts. It is worth noting that all students that participated in the follow-up interviews validated the concepts and categories presented in Table~\ref{tab:motivation_concepts}. As S\textsubscript{9} representatively said at the end of the interview: \hyphenquote{USenglish}{\textit{Yeah, yeah, I mean, I can see myself interested in many of these points [the categories] right, I did it [GSoC] for most of them}.}

\subsubsection{Career building}
Approximately 44\% of the students considered adding the GSoC experience to CV essential (see Q1 in Figures~\ref{fig:sts:motivators_assessment} and \ref{fig:sts:graph-venn-141-agreers}), preferring not to participate otherwise. Aside from \textit{technical challenge}, \textit{career building} was the motivation factor students were the least divided about, with $\approx$20\% of them being neutral on whether it was essential. Figure~\ref{fig:sts:graph-venn-141-agreers} (a) depicts that the students motivated by \textit{career building} were also mostly motivated by \textit{technical challenge} (84\%) followed by \textit{contribution to OSS} (58\%). Figure~\ref{fig:sts:graph-venn-141-agreers} (b) reveals that only one student was purely motivated by \textit{career building}.

We also analyzed students' textual answers to obtain additional information, which resulted in the concepts and categories shown in Table~\ref{tab:motivation_concepts} (see \textit{career building}). The analysis revealed, though not exclusively, that the students who mentioned the career as a motive for participation (27\%) mostly entered the program because \concept{GSoC would look good on their CVs} (31). Examples include S\textsubscript{79}: \hyphenquote{USenglish}{\textit{(...) adding the `Google' keyword on a resume was a good plus}}; and S\textsubscript{106}: \hyphenquote{USenglish}{\textit{I needed some real experience to my CV}.}

While a few other students considered \textit{career building} (7) to be among their primary motivation, their mentions were only vague, as per S\textsubscript{39}: \hyphenquote{USenglish}{\textit{I participated [in GSoC] because it was a great opportunity for my career}.} Moreover, \textbf{career building} (38) was a concern for several interviewees who declared they would not have given it up (5), revealing that their careers would still benefit from the: \concept{real-world development experience} (3); and \concept{interacting with OSS project members} (2).

\subsubsection{Contribution to OSS}

The students who explicitly stated to have entered GSoC motivated by contributing to OSS were grouped into the \textbf{contribution to OSS} (81) category (see Table~\ref{tab:motivation_concepts}).

Some students mentioned being driven by the \concept{GSoC/OSS project itself} (8), such as S\textsubscript{136}: \hyphenquote{USenglish}{\textit{I wanted to add a feature to an open source media player, and I felt like GSoC would motivate me to implement this feature in a short amount of time;}} and S\textsubscript{85}: \hyphenquote{USenglish}{\textit{I was interested in contributing to Free/Open source libraries and trying something new}.} The students did not mention they were interested in becoming frequent contributors.

We found cases of students who entered GSoC motivated by the \concept{OSS culture and philosophy} (16), such as S\textsubscript{73} who said: \hyphenquote{USenglish}{\textit{I'm passionate about FOSS and all philosophy around it};} S\textsubscript{58}: \hyphenquote{USenglish}{\textit{I was always attracted to the idea of contributing code for good};} and S\textsubscript{11}: \hyphenquote{USenglish}{\textit{I love coding and the idea of contributions to open source and helping others is too good}.}

Several OSS projects are known to have high entry barriers for newcomers~\cite{Steinmacher2014}, and in some cases, students considered that \concept{GSoC lowers entry barriers} (9), such as S\textsubscript{135}: \hyphenquote{USenglish}{\textit{I wanted to get involved developing OSS but found there to be a high barrier to entry (...) The goal for me was primarily to help break into the OSS community, which felt difficult to penetrate at the time}.}More often, students considered GSoC an opportunity to \concept{interact with OSS mentor or other community members} (21), such as S\textsubscript{48}, who said: \hyphenquote{USenglish}{\textit{It was a chance to interact with an OSS community}.} Although most students were not contributors to the GSoC projects before kickoff (see Table~\ref{tab:sts:contrib_2_GSoC}), a significant minority (44\%) had already contributed. Besides, most of the students reported having some previous experience in contributing to OSS projects (see Table~\ref{tab:sts:contrib_2_OSS}).

We also found students (2) that engaged in OSS projects to increase their odds of participating in GSoC. As evidenced by S\textsubscript{3}: \hyphenquote{USenglish}{\textit{I knew I had to do GSoC for which I started contributing to FOSS}.} This confirms what we found in students' and mentors' blogs\footnote{\href{https://danielpocock.com/getting-selected-for-google-summer-of-code-2016}{ https://danielpocock.com/getting-selected-for-google-summer-of-code-2016}} with tips on how to be accepted, suggesting that the candidates get involved with the community to increase their chances. We also found this advice in community wikis: \hyphenquote{USenglish}{\textit{Previous contributions to Octave are a condition for acceptance. In this way, we hope to select students who are familiar with the codebase and able to start their project quickly}.}\footnote{\href{https://wiki.octave.org/GSoC_2018_application}{ https://wiki.octave.org/GSoC\_2018\_application}}
Other strategy employed by students (2) was to select projects that few other students would be likely interested.

\begin{figure}
  \centering
  \includegraphics[scale=0.3]{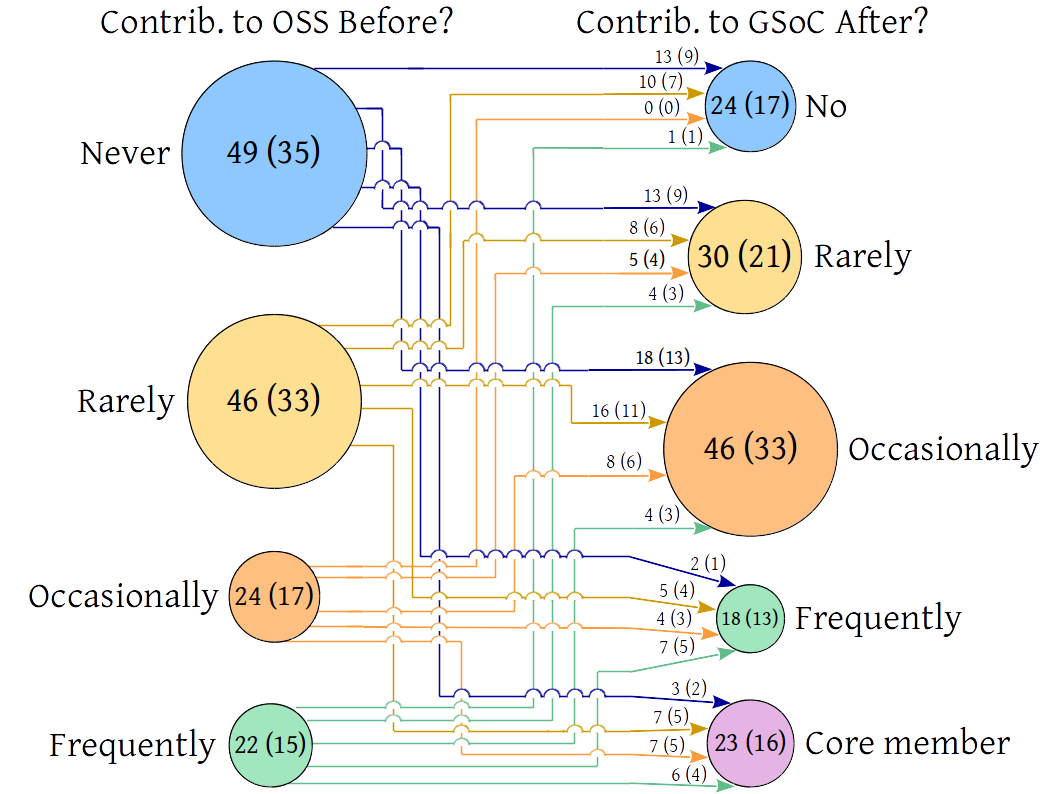}
  \caption{Contribution frequency to OSS Before and to the GSoC projects After the program. Students' count (\%).}
  \label{fig:sts:contrib-2-oss-bef-2-gsoc-aft}
\end{figure}

Figure~\ref{fig:sts:contrib-2-oss-bef-2-gsoc-aft} illustrates the relationship between the self-reported contribution frequency to OSS projects before kickoff and the assigned GSoC projects after the program. We can observe that 75 students ($\approx$53\%) reported an increase in contribution frequencies after GSoC. The 29 students ($\approx$21\%) who before GSoC had occasionally (at most) contributed to OSS projects remained as such after the program concerning contributions to the GSoC projects. Also, the 13 students ($\approx$9\%) who self-reported to be frequent contributors to OSS projects before the program remained as such after the program concerning contributions to GSoC projects. In contrast, 24 students ($\approx$17\%) lowered their contributions to GSoC projects compared to how frequently they contributed to OSS projects before the program's kickoff.

Contributing to OSS projects was ranked as the second most essential motivator (see Figure \ref{fig:sts:motivators_assessment}a), which is also confirmed by the students' responses coding (see Table \ref{tab:motivation_concepts}). In addition, most students entered GSoC with intentions to keep contributing (\textit{'Yes'} and \textit{'Definitely'}, which totals $\approx$57\%) (see Table~\ref{tab:sts:contrib_intention}). Together, these results suggest high retention rates. However, we interpret (and moderate) these results in light of our previous quantitative study \cite{Silva2017c}, which revealed that only a fraction ($\approx$16\%) of the students kept contributing after a few months. In this sense, this research confirms the work of Roberts et al. \cite{Roberts2006a}, who found in a longitudinal study that initial developers' motivations did not translated into increased retention. Nevertheless, both this research and our previous work \cite{Silva2017c} suggest a small group of students which indeed became frequent developers.

\begin{tabular}{cc}
    \begin{minipage}{.5\linewidth}
  \scriptsize
  \captionof{table}{Before GSoC, did you contribute to the project you've chosen for the program?}
  \label{tab:sts:contrib_2_GSoC}
    \begin{tabular}{lr}
    \hline\noalign{\smallskip}
      \textbf{Responses} & \textbf{Count (\%)}          \\
    \noalign{\smallskip}\hline\noalign{\smallskip}
      Never                         & 79 (56.0)    \\
      Rarely                        & 19 (13.5)    \\
      Occasionally                  & 10 \hspace{0.15cm}(7.1)    \\
      Frequently                    & 14 \hspace{0.15cm}(9.9)    \\
      My project started in GSoC    & 13 \hspace{0.15cm}(9.2)     \\
      Core member                   &  6 \hspace{0.15cm}(4.3)     \\
    \noalign{\smallskip}\hline
  \end{tabular}
\end{minipage}&

\begin{minipage}{.5\linewidth}
  \centering
    \scriptsize
  \captionof{table}{Before GSoC, did you contribute to OSS projects other than your own?}
  \label{tab:sts:contrib_2_OSS}
  \begin{tabular}{lr}
    \hline\noalign{\smallskip}
      \textbf{Responses}          & \textbf{Count (\%)}         \\
    \noalign{\smallskip}\hline\noalign{\smallskip}
      Never                       & 49 (34.7)        \\
      Rarely                      & 46 (32.6)        \\
      Occasionally                & 24 (17.0)        \\
      Frequently                  & 22 (15.6)        \\
    \noalign{\smallskip}\hline
  \end{tabular}
\end{minipage}

\\
\\
  \begin{minipage}{.5\linewidth}
      \centering
      \scriptsize
      \captionof{table}{Before GSoC, did you intend to continue contributing to the project?}
      \label{tab:sts:contrib_intention}
      \begin{tabular}{lr}
        \hline\noalign{\smallskip}
          \textbf{Responses} & \textbf{Count (\%)}          \\
        \noalign{\smallskip}\hline\noalign{\smallskip}
          Not at all         &  8 \hspace{0.05cm} (5.7)    \\
          No                 & 11 \hspace{0.05cm} (7.8)    \\
          Maybe              & 42 (29.8)    \\
          Yes                & 40 (28.4)    \\
          Definitely yes     & 40 (28.4)   \\
        \noalign{\smallskip}\hline
      \end{tabular}
  \end{minipage}&

  \begin{minipage}{.5\linewidth}
      \centering
      \scriptsize
       \captionof{table}{Have you actually continued contributing?}
      \label{tab:sts:actually_contributed}
      \begin{tabular}{lr}
        \hline\noalign{\smallskip}
          \textbf{Responses} & \textbf{Count (\%)}          \\
        \noalign{\smallskip}\hline\noalign{\smallskip}
          No               & 24 (17.0)    \\
          Rarely           & 30 (21.3)    \\
          Occasionally     & 46 (32.6)    \\
          Frequently       & 18 (12.8)   \\
          Core member      & 23 (16.3)   \\
        \noalign{\smallskip}\hline
      \end{tabular}
  \end{minipage}
\end{tabular}

\subsubsection{Peer recognition}
Only a quarter of the students ($\approx$25\%) considered \textit{peer recognition} essential for participation (see Q3 in Figure~\ref{fig:sts:motivators_assessment} and Figure~\ref{fig:sts:graph-venn-141-agreers}).

Often, students referred to \textit{peer recognition} concerning \concept{prestige} (5) of the program among their peers of yet \concept{bragging rights} (4).

\subsubsection{Stipends}
Around 30\% of the students considered \textit{stipends} essential for participating in GSoC, even though this motivation factor had the largest number of neutral students (see Figure~\ref{fig:sts:motivators_assessment} and Figure~\ref{fig:sts:graph-venn-141-agreers}).

Some students revealed the roles the \textit{stipends} played. In several cases, students used the \textit{stipends} for the \concept{payment of their tuition} (13).

Often, the \textit{stipends} were used as a \concept{source of funding} (13). We used this concept when the \textit{stipends} were used for \concept{living expenses} (10), as a means to make students' participation feasible, such as explained by S\textsubscript{115}: \hyphenquote{USenglish}{\textit{As a student[,] I need to earn money for existence}}, and S\textsubscript{125}: \hyphenquote{USenglish}{\textit{I needed the stipend for living expenses}.}

During the interviews, we found that students used the \textit{stipends} to \concept{buy hardware equipment} (1), coded as \concept{source of funding} (13). As S\textsubscript{47} said: \hyphenquote{USenglish}{\textit{I used that [the stipends] to purchase hardware equipment so I could improve my development environment}.} Furthermore, we considered \concept{source of funding} (13) when existing project members could \concept{dedicate time and efforts to their projects} (2), such as S\textsubscript{6}: \hyphenquote{USenglish}{\textit{I was already contributing to the OSS project before the GSoC although that was in my free time. GSoC was a chance to really spend time for the project}}; and S\textsubscript{111}: \hyphenquote{USenglish}{\textit{GSoC was a chance for us to have a core member work on the project full time instead of just in the spare time and this helped to get lots of development and some crucial refactoring done}.}

Alternatively, some other students viewed stipends as compensation for either the service provided or the time spent, which we labeled as \concept{stipends as compensation} (10), such as explained by S\textsubscript{40}: \hyphenquote{USenglish}{\textit{I would prefer to get paid for my time. Otherwise[, I would have] contributed to open source without GSoC}}; and S\textsubscript{86}: \hyphenquote{USenglish}{\textit{I like to be paid for my work}.}

Many responses mentioned the stipends to be significant, such as S\textsubscript{84}, who commented: \hyphenquote{USenglish}{\textit{It was a really cool opportunity to (...) get a (huge) amount of money (...)}.} Since the stipends' role was not explicitly stated, we present these counts in the same line as the category. This rationale also was applied to students who were motivated by \concept{currency conversion} (2) rates, such as S\textsubscript{137}, who said: \hyphenquote{USenglish}{\textit{For the financial incentive (which is quite a big amount in my country) and for the opportunity to contribute to OSS projects}.} These students resided at Sri Lanka and Belarus when they participated in GSoC.

Stipend-motivated participation incited different sentiments in the students. Although most students' responses were neutral (120) towards the stipends, some responses had a positive tone (8), typically linking the payments to the heart of the program. As S\textsubscript{95} answered when asked if he would enter a no-stipend hypothetical-GSoC: \hyphenquote{USenglish}{\textit{That's a weird question, the point of GSoC is the stipend, [otherwise] there wouldn't be any GSoC}.} On the other hand, we also identified a minority of students (3) with negative sentiments towards participation motivated by payments. As S\textsubscript{52} mentioned: \hyphenquote{USenglish}{\textit{There are many people who try GSoC merely for the money! That's something of utter shame. People should contribute only if they're genuinely interested and not for the money}.}

\subsubsection{Learning}
Several students reported that the potential \textbf{learning} (58) experience provided by GSoC was among their motivation for participation, mostly for the \concept{real-world development experience} (51), which means that the students wanted to improve their programming skills or be introduced to software engineering practices. As S\textsubscript{67} detailed: \hyphenquote{USenglish}{\textit{I was looking for an internship/summer experience and GSoC caught my eye because it seems like a good way to improve programming skills (...)}.}

We also found evidence of some students motivated to enter GSoC because they wanted to \concept{gain other skills} (2) (other than programming), such as S\textsubscript{99}, who described his interest: \hyphenquote{USenglish}{\textit{To improve English}.} In addition, a few students vaguely mentioned \textit{learning} (5), without specifying what they wanted to learn.

\subsubsection{Academic}

While a few students vaguely reported participating in GSoC for \textit{academic} (7) concerns, others wanted an alternative to traditional \concept{internships} (6). These students often indicated as a primary motivating factor the flexibility that GSoC offered, such as working remotely. The quote of S\textsubscript{109} exemplifies these cases: \hyphenquote{USenglish}{\textit{It was a good summer internship, getting good internship locally was difficult for me}.} The work conditions offered by GSoC motivated another student. As S\textsubscript{118} explained his interest: \hyphenquote{USenglish}{\textit{[I] needed a [low-pressure] internship like this}.}

Similarly, other students driven by \textit{academic} motives mentioned the need for the accomplishment of \concept{summer projects} (9). As S\textsubscript{58} said: \hyphenquote{USenglish}{\textit{I was looking for a summer project}.} Due to the similarity, we grouped the concepts \concept{internships} (6) and \concept{summer projects} (9) into a single \concept{internships/summer projects} (15) concept. Also, graduate students mentioned participating in the program for \concept{research purposes} (4), such as S\textsubscript{130}, who commented: \hyphenquote{USenglish}{\textit{I was a graduate student looking for summer funding and I wanted to improve my coding for my research}.}

During the interview, two students added that the participation in GSoC could be used for obtaining \concept{course credits} (2) in their college. As S\textsubscript{5} said: \hyphenquote{USenglish}{\textit{There are some students I know that specifically did GSoC just for the college course credit}.}

\subsubsection{Technical challenge}
Approximately 67\% of the students considered \textit{technical challenge} essential for participation (see Q5 in Figure~\ref{fig:sts:motivators_assessment} and Figure~\ref{fig:sts:graph-venn-141-agreers}). It was the motivation factor for which the largest number of students declared they would not enter GSoC without and that the students were least divided.

Surprisingly, analyzing our coding, we found that \concept{technical challenge} (5) was the least mentioned motivation factor (see Table~\ref{tab:motivation_concepts}), with only a few mentions. Still, these mentions were subtle. For instance, S\textsubscript{72} said: \hyphenquote{USenglish}{\textit{It's challenging, it's interesting, and it's [paid]}.}

\takeaway{Answer for RQ1:}{ Based on our data, the students typically entered GSoC for a paid experience in which they could use the practical knowledge obtained from participation for building their career portfolio. Nevertheless, some students entered mainly to be able to contribute to OSS projects.}

Although it is not the focus of this research to investigate differences in students' motivation by gender, country of residence, and education level, we offer some analysis under these perspectives.  Our sample indicates that GSoC is male-oriented (as with the broader software engineering field) and our data is insufficient for segmenting by gender.  We did not find significant differences in students' motivation when we grouped the countries of residence by development level. Finally,  career-driven participations seems correlated with an age group (21-25). Additional research is necessary to understand and validate these differences.


\subsection{Students' Motivations From Mentors' Perspective (RQ2)}

\begin{figure}[tb]
  \centering
  \includegraphics[scale=1.1]{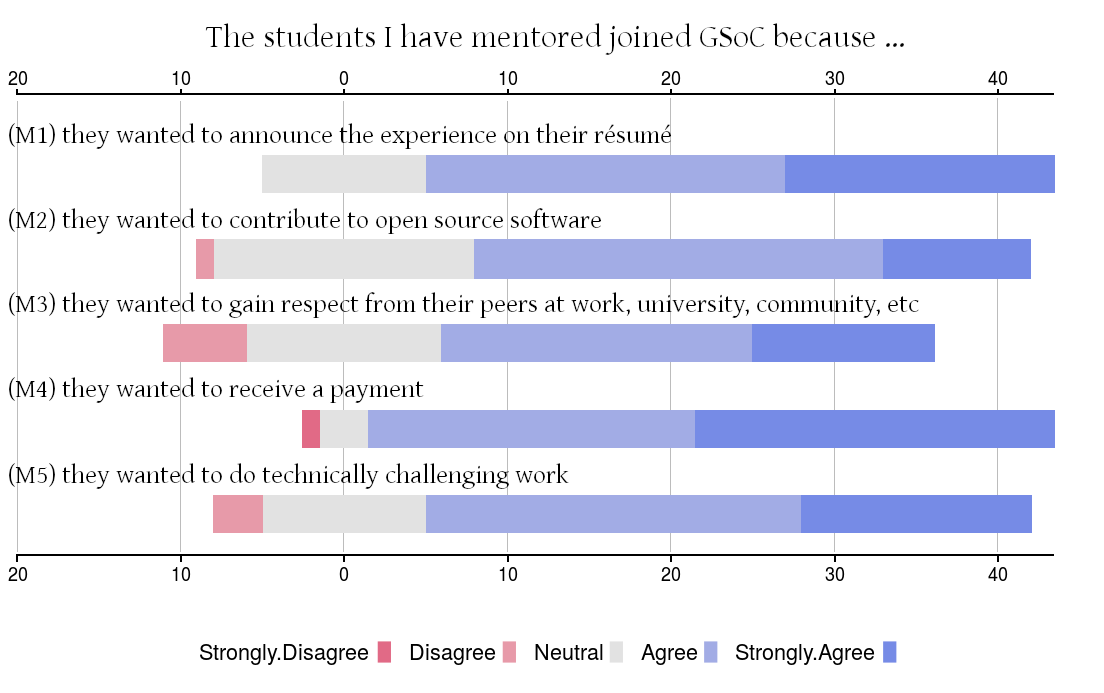}
  \caption{Mentors' perception on the students' motivation for entering GSoC}
  \label{fig:mnt:motivators_assessment}
\end{figure}

\begin{figure}
  \includegraphics[trim=2cm 17.5cm 2.2cm 2.5cm, clip, width=\textwidth]{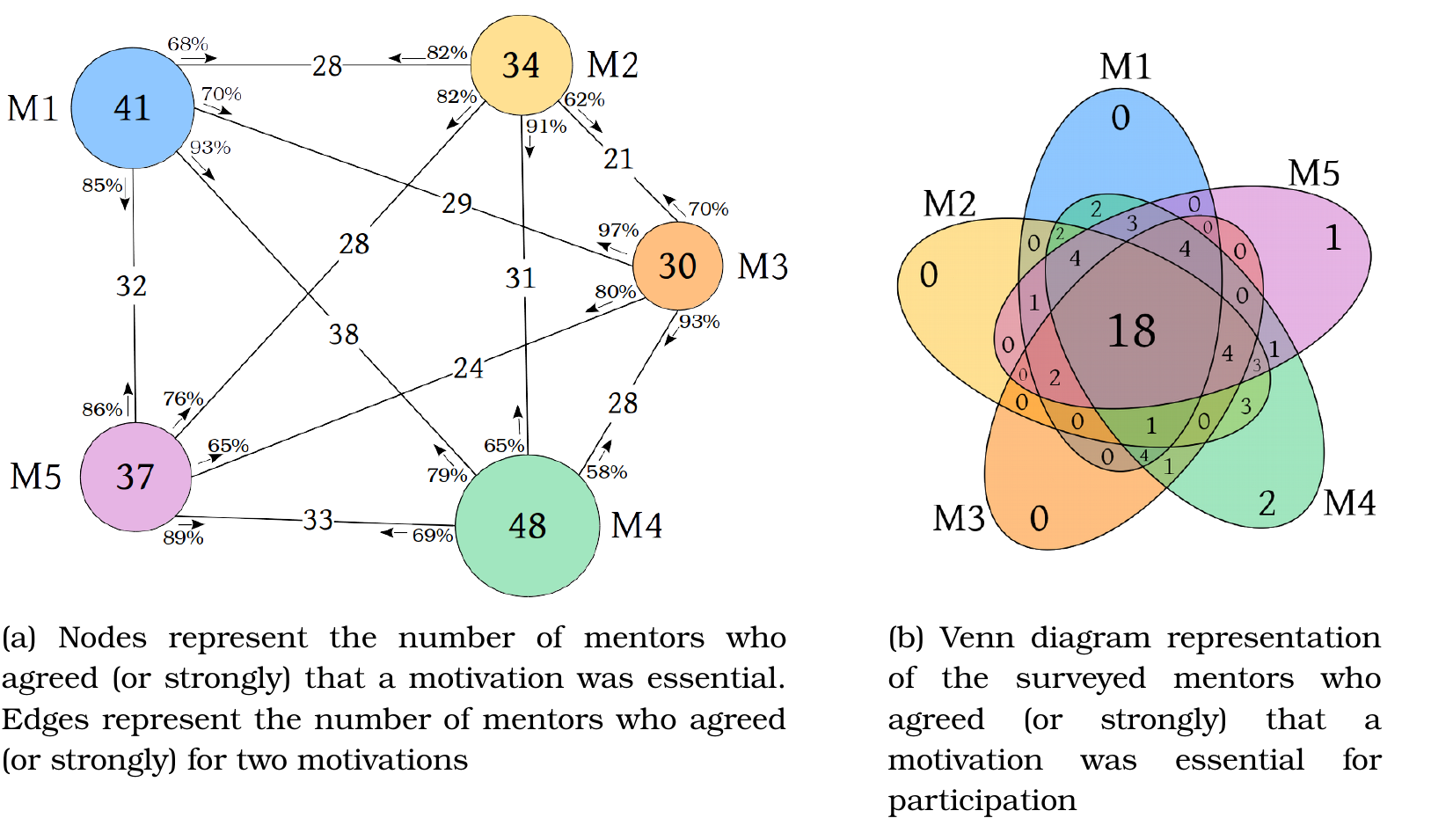}
  \caption{Count of students' motivation according to mentors in a graph (a) and in a Venn diagram (b). Career building (M1); contribution to OSS (M2); peer recognition (M3); stipends (M4); technical challenge (M5)}
  \label{fig:mnt:graph-venn-agreers}
\end{figure}

Figure~\ref{fig:mnt:motivators_assessment} depicts in stacked bars the mentors' assessment on how essential the investigated motivation factors were for students to join GSoC. Similarly to Figure~\ref{fig:sts:graph-venn-141-agreers}, Figure~\ref{fig:mnt:graph-venn-agreers} offers additional perspectives.

\subsubsection{Career building}
Approximately 77\% of mentors agreed that students entered GSoC for announcing the experience in their CV (see M1 in Figure~\ref{fig:mnt:motivators_assessment} and Figure~\ref{fig:mnt:graph-venn-agreers}). It is worth noting that \textit{career building} was the only motivating factor for which no mentor disagreed that it was essential for students.

In Figure~\ref{fig:mnt:graph-venn-agreers} (a), we can observe that virtually all the mentors who agreed that \textit{career building} was essential (M1, edge: 93\%) also agreed that \textit{stipends} were essential (M4). The remaining edges equally show that more than 2/3 of the mentors in M1 also considered the remaining motivation factors essential for participation. Figure~\ref{fig:mnt:graph-venn-agreers} (b) shows that no mentor considered that students were only trying to improve their CVs by participating in GSoC. Instead, mentors tended to assess students' motivations as being multifaceted, to the point that approximately 1/3 of the mentors (i.e., 18 mentors) considered all motivation factors essential for participation.

In the answers to our open-ended questions, some mentors mentioned \concept{CV improvement} (9) as a motive for students to enter GSoC. As M\textsubscript{36} representatively said: \hyphenquote{USenglish}{\textit{They [the students] are interested in building their CV, being recognized as part of a Google's program}.}

\subsubsection{Contribution to OSS}
Around 64\% of mentors agreed that students joined GSoC motivated by the \textit{contribution to OSS} (see M2 in Figure~\ref{fig:mnt:motivators_assessment} and Figure~\ref{fig:mnt:graph-venn-agreers}). While \textit{contribution to OSS} was the second most essential motivation factor in the students' perception, mentors' assessment was that \textit{contribution to OSS} is the second least essential factor (compare Q2 in Figure~\ref{fig:sts:motivators_assessment} to M2 in Figure~\ref{fig:mnt:motivators_assessment}).

In general, mentors perceived students as contributors to OSS projects (see Table~\ref{tab:mnt:contrib_tab} (a) and (b)), though in several cases mentors classified contribution frequency as rare. This perception may explain why mentors possibly underestimated (compared to the other motivation factors) how essential \textit{contribution to OSS} was for the students since in mentors' view most students already had that experience.

We also found potential disparities among mentors' and students' perception regarding contributing to OSS before GSoC. In Table~\ref{tab:mnt:contrib_tab} (a), we can observe that $\approx$13\% of the mentors in our sample considered that students had never contributed to OSS, while $\approx$35\% of the students self-reported to have never contributed to OSS before GSoC. On the other hand, while $\approx$3\% of the mentors reported that students were frequent contributors before GSoC (see Table~\ref{tab:mnt:contrib_tab} (a)), 16\% of the students self-reported to be frequent contributors (compare to Table~\ref{tab:sts:contrib_2_OSS}). A similar disparity occurs when we compare the students' (Table~\ref{tab:sts:contrib_2_GSoC} and mentors' (Table~\ref{tab:mnt:contrib_tab} (b)) perception of the frequency of previous contributions to GSoC projects.

These disparities can be in part explained considering that the students that mentors referred to were not necessarily GSoC first-timers; were active project contributors before GSoC, and started contributing to OSS/GSoC projects to increase the odds of being accepted in GSoC. Another possible explanation is that students' and mentors' view differed towards what they considered to be a frequent contributor.

\begin{table}
  \centering
  \footnotesize
  \caption{(a) In your experience, how often were your GSoC students contributors to OSS software projects (other than their own) before the program? \\ (b) Were they already contributors to the project you mentored before GSoC? \\ (c) How often do students keep contributing to the projects you mentored after the program? }

  \label{tab:mnt:contrib_tab}
    \includegraphics[trim=2cm 22.6cm 11cm 2.5cm, clip, scale=0.8]{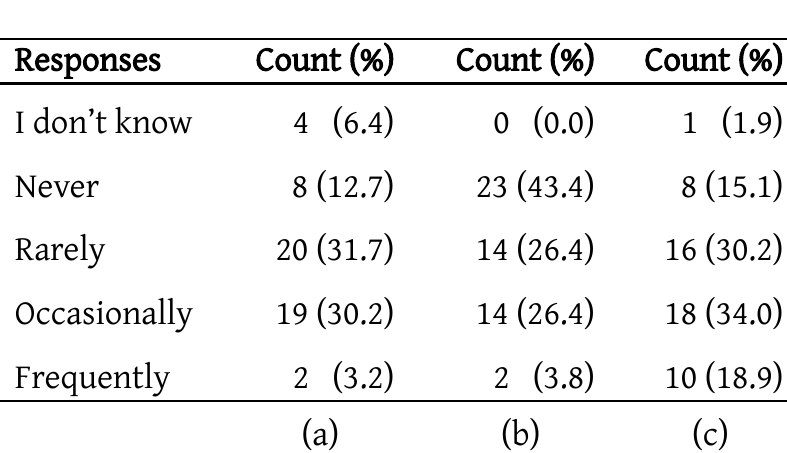}
\end{table}

Figure~\ref{fig:mnt:graph-venn-agreers} show that mentors perceived a strong link between the \textit{contribution to OSS} and \textit{stipends} motivation factors. We can observe that 91\% of the mentors who considered \textit{contribution to OSS} an essential motivation factor did the same for \textit{stipends} (see M2 in Figure~\ref{fig:mnt:graph-venn-agreers}a). The remaining motivation factors also had more than 2/3 of the mentors who considered them essential, except for \textit{peer recognition} (M3).

The coding of mentors' answers revealed that \concept{interaction with OSS community members} (5) is a primary interest, even though there was a subtle mention to the \concept{OSS project itself} (1) as a motive. We also found evidence that the GSoC selection process can potentially make candidates contribute to OSS projects as a means to get accepted in the program (1).

\subsubsection{Peer recognition}
Around 57\% of mentors considered \textit{peer recognition} an essential motivation factor for students, being the least essential when compared to the other studied factors (see M3 in Figure~\ref{fig:mnt:motivators_assessment} and Figure~\ref{fig:mnt:graph-venn-agreers}). This finding is coherent with students' assessment, which also considered \textit{peer recognition} the least essential motivation factor.

In Figure~\ref{fig:mnt:graph-venn-agreers} (a), we can observe that virtually every mentor who considered \textit{peer recognition} essential also did the same for \textit{career building} (see M3, edge: 97\%) and \textit{stipends} (see M3, edge: 93\%), although more than 2/3 of mentors considered the other motivation factors essential.

In their textual answers, mentors rarely mentioned \textbf{peer recognition} (2) as a motive for participating in GSoC, and we only found two subtle mentions. M\textsubscript{15}: \hyphenquote{USenglish}{\textit{Kudos and getting paid}} and M\textsubscript{27}, who was more specific: \hyphenquote{USenglish}{\textit{...for bragging rights}.}

\subsubsection{Stipends}
According to mentors, the \textit{stipends} were an essential motivation factor for students (see M4 in Figure~\ref{fig:mnt:motivators_assessment} and Figure~\ref{fig:mnt:graph-venn-agreers}), with a consensus of $\approx$91\%. We can see in Figure~\ref{fig:mnt:graph-venn-agreers} (a) that most mentors classified students' motivation as a combination of \textit{stipends} and other factors, typically \textit{career building} ($\approx$79\% of cases). In Figure~\ref{fig:mnt:graph-venn-agreers} (b), we can observe that two mentors judged that the \textit{stipends} alone sufficed for students to enter GSoC.

The coding of mentors' answers was consistent with the previous finding, showing that the \textbf{stipends} (24) were the most cited motivation factor for participation (see Table~\ref{tab:motivation_concepts}), even though often the mentors mentioned the \textit{stipends} (21) broadly, without offering any context.

Nevertheless, a few mentors mentioned stipends as a \concept{source of funding} (2). For instance, when M\textsubscript{40} commented on what his students were most interested in by entering GSoC: \hyphenquote{USenglish}{\textit{Money. Honestly, they're students, which I'm pretty sure is a synonym for starving and broke}.} We also could find evidence for \concept{currency conversion} (1) as a motive for participation. For example, M\textsubscript{10}, who said: \hyphenquote{USenglish}{\textit{The money seems to be a strong incentive. Especially in countries where approx \$5,500 USD carries a lot of purchasing power}.} (in most GSoC editions so far, the stipends were not proportional to purchasing power). No mentor mentioned \concept{stipends as compensation} (0) as a motive.

Additionally, while several mentors who commented on \textit{stipends} as a motive implied a neutral (30) or positive (1) tone in their answers, some mentors (3) indicated a negative tone. As M\textsubscript{2} said: \hyphenquote{USenglish}{\textit{Sadly, the money}}; and M\textsubscript{46}: \hyphenquote{USenglish}{\textit{I guess good students are more interested in learning and contributing, and not so good students by improving their CV and money}}; and M\textsubscript{33}, who commented: \hyphenquote{USenglish}{\textit{Many of the students I have mentored (15 or so at this point?) seemed to want to do the bare minimum to pass their deadlines and get paid}.} Encouragingly, we found evidence of mentors with a different experience. As M\textsubscript{11} said: \hyphenquote{USenglish}{\textit{Money is a strong motivator to join the program obviously, but most of them continue contributing after that factor disappears}.}

\subsubsection{Technical challenge}
Approximately 70\% of mentors agreed that the \concept{technical challenge} (2) that the GSoC projects placed on their students is something the students had aimed for (see M4 in Figure~\ref{fig:mnt:motivators_assessment}). However, as with the students' answers, the \concept{technical challenge} (2) motivation factor had few mentions in mentors' coding.

\subsubsection{Academic}
Several mentors mentioned that \textbf{academic} (6) concerns motivated students to enter GSoC. Except for a single generic mention to \textit{academic} (1) as a motivation factor, mentors identified that their students entered GSoC for \concept{course credits} (1), for \concept{research purposes} (4), and \concept{internship/summer projects} (4).

\subsubsection{Learning}
Several mentors commented that \textbf{learning} (17) plays a central role in motivating students to enter GSoC. Only a few mentors mentioned \textit{learning} (4) broadly. More commonly, mentors linked \textit{learning} to the acquiring of \concept{real-world development experience} (13).

\takeaway{Answer for RQ2: }{Mentors in our sample perceive their students as entering GSoC for the technical learning, in a favorable environment, which the mentors portrayed as including stipends and mentoring, mainly for building the students' career portfolio.}


\section{Discussion}
\label{sec:discussion}

In this section, we review and discuss our findings. The literature on motivations to join OSS is mostly focused on contributors who are self-guided volunteers. In this research, we investigate whether the introduction of incentives offered by Summer of Code programs add new elements to the students' motivation.

(RQ1) Our research is the first to document what motivates students to participate in Summer of Code programs (Table~\ref{tab:motivation_concepts}).
Even if some of the factors are similar to the context in which OSS developers voluntarily contribute to OSS projects (see \cite{VonKrogh2012} for a review) the contribution the projects through Summer of Code context is quite different, leading to a different prioritization of factors. Additionally, three motivating factors seem to be new: participate in GSoC for \textit{taking advantage of currency conversion}; \textit{obtaining course credits}; and \textit{lowering OSS projects' entry barriers}.

(RQ2) We also document the mentors' perception of the students' motivations (see Table~\ref{tab:motivation_concepts}), which is also not targeted by previous research. Mentors provide a perspective that considers the project's point of view, the comparison to non-GSoC newcomers, and an external view of the students' motivation to enter Summer of Code programs. In essence, mentors perceived students' motivation as a pursuit of tangible rewards such as \textit{stipends}, and the \textit{learning} of \textit{technical skills} to be used mainly for \textit{career building}.

Regarding students' retention, our findings suggest that most students do not remain contributing to GSoC projects after the program, regardless of their initial intentions (see Table~\ref{tab:sts:contrib_intention}). This finding is supported by our previous work \cite{Silva2017c}, in which we found that most students stopped contributing after GSoC, while the students who remained had only a few commits to the GSoC projects. Encouragingly, as with the findings of this research (see Figure~\ref{fig:sts:motivators_assessment} and Figure~\ref{fig:sts:graph-venn-141-agreers}), our previous work \cite{Silva2017c} indicated that some students became frequent contributors after GSoC. Thus, it seems that most students enter the program for an enriching (work) experience that cannot be detached from the name of a high profile software company (such as Google). In this sense, our findings suggest that most OSS projects can expect feature development from participating in GSoC.

Furthermore, our findings suggest that students are reluctant to admit financial motivation according to mentors' answers.

Nevertheless, we could notice that for students with 2 to 5 years (61 students) of previous software development experience would still enter a hypothetical-GSoC that did not offer any stipends, as opposed to the ones with the same time experience who would not (20). In contrast, the students with 10 or more years (15 students) of prior development experience would not enter a hypothetical-GSoC with no payments, as opposed to the ones within the same experience range (5) who would still enter. Therefore, although the stipends is an important motivator, it seems to be essential for participation for students with high software development experience, while the students who lack development experience value participation in GSoC for boosting their careers.

Indeed, low retention levels (or high levels of absenteeism in some contexts) is the most expected outcome in volunteer engagement programs (see \cite{Smith2014} for the firefighting community in the USA;~\cite{Lacetera2013} for blood donation, and~\cite{Resnick2009,Zhu2013} for online communities). Encouragingly, regardless of their motivation for entering GSoC, students self-reported an increase of their previous contribution level to the assigned GSoC projects in $\approx$53\% of cases (see Figure~\ref{fig:sts:contrib-2-oss-bef-2-gsoc-aft}).

Nevertheless, low retention rates may be demotivating for some mentors, mainly because they invest a lot of effort and time into mentoring. As mentioned by a mentor: \hyphenquote{USenglish}{\textit{I participated in GSoC as a mentor (...) While it didn't 'cost' me anything in dollars, it cost me probably 200 hours of my time.}}\footnote{\href{https://mail-archives.apache.org/mod\_mbox/community-dev//201612.mbox/\%3C8a807ec4-af85-3f8d-d080-1bc30a872898@rcbowen.com\%3E}{https://mail-archives.apache.org/mod\_mbox/community-dev//201612.mbox/\%3C8a807ec4-af85-3f8d-d080-1bc30a872898@rcbowen.com\%3E}} High-quality mentoring is labor-intensive and time-consuming and, in several cases, offered by volunteer OSS project members. While offering dedicated mentorship plus designing a high-level Summer of Code project could potentially enrich students' experience in contributing to OSS projects, it may have the adverse effect of lowering mentors motivation. This seems to be a dilemma faced by the Debian community, which decided not to participate in GSoC 2017, as shown by the following excerpt from a notification email: \hyphenquote{USenglish}{\textit{Debian will not take part [in GSoC] this year. Some of our recurring mentors have shown some signs of 'GSoC fatigue,' (...) let's have a summer to ourselves to recover (...) and come back next year.}} As previous research has shown that mentors themselves also face barriers \cite{balali2018newcomers}, our findings may---to some degree---assist mentors by showing what aspects of GSoC the students are most interested in.

Our findings revealed that there are students whose primary goal was to participate in GSoC, not necessarily to contribute to OSS projects. We speculate that these students would not have contributed to OSS projects otherwise. In addition, we conjecture that Summer of Code programs can potentially assist students in overcoming several of the onboarding barriers reported by Steinmacher et al.~\cite{Steinmacher2015}, which can be investigated in future research.

Previous research reports positive associations between receiving stipends and participating in OSS projects~\cite{Roberts2006a}. However, we found that the goals among stipend-driven students can be different. While some students understand the stipend as compensation for a service, others need it for living expenses or buying hardware equipment. Our findings trigger some questions to future research to understand these associations at a finer-grained level.

\subsection{Implications}
We list some implications of this study for different stakeholders.

\paragraph{OSS Projects} OSS project members should moderate their expectations about gaining long-term contributors. Although GSoC increased participation in GSoC projects in general, our findings suggest that most OSS projects did not achieve long-term contributors. Our data indicate that the OSS projects should consider GSoC as an investment in students' experience, in exchange for software feature development. OSS projects should consider that most of the students in our sample intended to become frequent contributors and a significant minority were neutral (see Table~\ref{tab:sts:contrib_intention}). This intention signals that providing students with rewards (e.g., certificates of contribution) that are meaningful to their goals (e.g., career building) should increase retention (or at least participation) rates. An alternative is to reward the students with seals of contribution or certificates associated with software companies (which do not need to sponsor students), enabling them to add these to their resum\'{e}s. In addition, Trainer and colleagues \cite{Trainer2014b} reported that the development of strong ties between students and project members (especially mentors) are associated with long-term contribution. We conjecture that this scheme could also be used with applicants not accepted in GSoC. Furthermore, GSoC is very competitive from the students' perspective. Thus, OSS projects should leverage contributions by attracting newcomers before GSoC, which not only could result in more contributions but also give mentors more time to assess suitable candidates.

\paragraph{Students} Students who want to take part as Summers of Code participants can benefit from the results of this study in many ways. First, our results show that students are encouraged to get involved with the OSS projects before the selection process, so they can showcase their abilities and willingness, increasing their odds of being accepted. Second, we could observe that Summers of Code bring rewards to the participants that go beyond the stipends. Students see these programs as great opportunities to build a portfolio and trigger their career, as can be observed in Table~\ref{tab:motivation_concepts}. Participants from developing countries report that participating in a program like GSoC increases students' visibility when seeking a job in large corporations. In addition, some students consider participating in GSoC as a chance of networking, enabling them to interact with OSS contributors and with "top of field people," as shown in Table~\ref{tab:motivation_concepts}. Third, students consider Summer of Code programs a good and flexible internship. It enables, for example, students that cannot commute or need to help their families during summer break, to participate in internships.

\paragraph{Summers of Code organizers} It is crucial that the organizers observe and value career advancements, by, for example, easing the access to the participants' list and providing certificates, something similar to what GSoC does. While looking online for the participants' email addresses, we analyzed the students' professional social networks profiles and noted that they indeed list the participation at GSoC as job experience. We could observe that a great part of the students' motives is not related to the stipends (see Table~\ref{tab:motivation_concepts}). Therefore, existing and potential new programs could offer the students a chance to participate without receiving stipends. By doing that, the projects would benefit from more newcomers, and the students would benefit from non-monetary rewards that the program offers. Besides, since one of the motives reported by the students was networking, Summers of Code programs would consider organizing regional meetups, inviting project members and participants, so they have a chance to meet the regional project members in person. Lastly, one thing that needs reflection from the Summer of Code organizers side is that, as participants come from all over the world (see statistics for 2017\footnote{\href{https://developers.google.com/open-source/gsoc/resources/stats\#2017}{ https://developers.google.com/open-source/gsoc/resources/stats\#2017}}), organizing the program in different periods, or making the calendar more flexible, would benefit students from countries in which the three-month break is from December to February.

\paragraph{Universities} Universities can also benefit from our results. Although Google does not classify GSoC as an internship,\footnote{\href{https://developers.google.com/open-source/gsoc/faq}{https://developers.google.com/open-source/gsoc/faq}} \textit{we evidenced that some universities use students' participation in the program for validating course credits}. Thus, universities could use our results to provide incentives and support students to get into GSoC as a way to both help the students and contribute to OSS. The students would get coding experience in a real setting, and they would be exposed to real challenges. The experience of a GSoC student could potentially enrich the experience of other students. Besides, validating course credits would be especially interesting for universities away from major cities, in which the internship possibilities do not offer technical challenges to enable students to put what they learned in practice.

\paragraph{Research.} This work offers different opportunities for researchers to extend our findings.

\textit{Legitimate Peripheral Participation (LPP)}.  LPP is frequently used to explain how newcomers engage in OSS projects (communities of practice)~\cite{Fang2009b}. However, our data indicate that LPP does not precisely describe the engagement process in OSS in GSoC in at least two ways. First, LPP assumes that students and mentors share the same goals, which would be to become frequent contributors to OSS projects. However, our findings indicate that most of the students in our sample were not primarily motivated to become frequent contributors (see Table~\ref{tab:motivation_concepts}). Second, contributing to OSS through GSoC may change the engagement process described by LPP. In several instances, students did not start at the margin, by first observing experienced members. Instead, they were individually guided---and sponsored---to become contributors. Also, according to LPP, by successfully contributing peripheral tasks, apprentices should be gradually legitimized by experienced members. Rather, the student-OSS-project relationship in a Summer-of-Code context is mediated by a contract. Thus, Summer-of-Code students have the time to dedicate themselves to the GSoC project, which provides them with an opportunity to develop strong social ties to mentors. Nevertheless, it is not clear from our data if relationships mediated by contracts could, in fact, legitimize students. Therefore, our findings indicate that more research is necessary to understand how students can be legitimized as project members in a Summer of Code context.

\textit{Self-Determination Theory (SDT)}. Deci and Ryan \cite{Deci1999a} suggested that an understanding of the effects of (participation) rewards requires a consideration of how the recipients (students) are likely to interpret the rewards. In particular, this interpretation is directly linked to the feelings of self-determination (autonomy) and competence (self-efficacy), which may affect intrinsic motivation. Even though we found that students' motivation comprises multiple dimensions, no research has focused on the effects of the rewards on intrinsic motivation, which several researchers consider essential in OSS context (e.g., \cite{Lakhani2005,Roberts2006a,Lakhani2003}).

\textit{Mentors}. Alternatively, we observed only students' motivation. However, to the best of our knowledge, mentors' motivation remains understudied. Understanding what drives mentors to support newcomers could benefit OSS projects and newcomers. Furthermore, it would be interesting to create an array of strategies that mentors use to deal with common problems such as candidates' selection, project creation, mentoring guidelines, and others.

\textit{Demographics}. Additionally, researchers could study students' demographics and how (or whether) potential differences influence students' motivation and contribution. Also, additional research is necessary to understand how companies see the participation in Summers of Code in their hiring processes.


\section{Limitations}
\label{sec:limitations}

This research has limitations, as described in the following.

\paragraph{Internal validity}
Surveys are typically subject to \textit{sampling} bias, namely \textit{self-selection} bias, which could distort our sample towards the students and mentors who chose to participate. Also, our sample of students and mentors is not sufficiently large for statistically grounded inferences. These threats could result in a biased sample, in which case it would not be representative of the actual population of students and mentors. Nevertheless, our focus is not on understanding how generalizable the motivation factors we found are but on identifying them.

Also, \textit{social desirability} can affect our data. For example, our data include negative viewpoints of students towards stipend-driven participation, which could indicate that a more significant number of students can perceive this factor as undesirable, underreporting (consciously or not) how essential the stipends were for their engagement.

Another threat is the data classifications' subjectivity. We used coding procedures to mitigate this threat, given that our findings are grounded in the data collected. Additionally, we discussed the analysis process, codes, concepts, categories, and the findings among the authors to promote a better validation of the interpretations through agreement. Moreover, the data collected via Likert-scale in the survey and follow-up interviews confirmed our coding scheme.

\paragraph{External validity} The main limitation affecting external validity is our focus on GSoC. Also, we only investigated the GSoC editions from 2010 to 2015. Furthermore, as few respondents identified themselves as female or other, our results may be biased towards male students. Although we are confident that most of our results are also valid in other settings, we leave this investigation to future research.

\section{Conclusion}
\label{sec:conclusion}

In this paper, we investigated what motivates students to participate in Google Summer of Code (GSoC). More specifically, we surveyed 141 students and 53 mentors that participated in different GSoC editions, followed by ten confirmatory interviews.

Our findings suggest that students typically participate in GSoC for work experience, rather than becoming a frequent OSS contributor. We also revealed that the students considered essential for participation: \textit{technical challenge}, \textit{contributing to OSS}, \textit{build their careers}, \textit{stipends},  \textit{peer recognition}, \textit{Learning}, and \textit{academic} concerns. From the mentors' perspective, students' motivation is mostly related to tangible rewards, such as stipends and technical learning to be used for career building.

In general, we found that participation in Summers of Code provided some OSS projects with new collaborators, even though this is not the typical scenario. OSS projects can use our findings to design strategies to increase attractiveness and retention.

\section{Acknowledgement}

This work is partially supported by the CNPq (430642/2016-4); FAPESP (Grant 2015/24527-3);
and the National Science Foundation (Grant numbers 1815503 and 1900903)

\bibliography{jss}

\end{document}